\documentclass[aps,pla,twocolumn]{revtex4}
\usepackage{amssymb}
\usepackage{graphicx}
\usepackage{amsmath}
\usepackage{dcolumn}
\usepackage{bm}

\newcommand{\beq}{\begin{equation}}
\newcommand{\eeq}{\end{equation}}
\newcommand{\beqa}{\begin{eqnarray}}
\newcommand{\eeqa}{\end{eqnarray}}

\def\oc#1{{ Opt.\ Commun.} {\bf#1}}

\def\jpb#1{{ J.\ Phys.\ B} {\bf#1}}
\def\jpa#1{{ J.\ Phys.\ A} {\bf#1}}

\def\pra#1{{ Phys.\ Rev. A\/} {\bf#1}}
\def\prb#1{{ Phys.\ Rev. B\/} {\bf#1}}

\def\pre#1{{ Phys.\ Rev. E\/} {\bf#1}}
\def\prl#1{{ Phys.\ Rev.\ Lett.} {\bf#1}}
\def\qic#1{{ Quant.\ Inf.\ Comp.} {\bf#1}}
\def\sci#1{{ Science} {\bf#1}}

\begin{document}

\title{Initial Conditions and Entanglement Sudden Death}
\author{Xiao-Feng Qian}
\author{J.H. Eberly}
\affiliation{Rochester Theory Center and Department of Physics \& Astronomy\\
University of Rochester, Rochester, New York 14627}
\date{\today }

\begin{abstract}
We report results bearing on the behavior of non-local decoherence
and its potential for being managed or even controlled. The
decoherence process known as entanglement sudden death (ESD) can
drive prepared entanglement to zero at the same time that local
coherences and fidelity remain non-zero. For a generic
ESD-susceptible Bell superposition state, we provide rules
restricting the occurrence and timing of ESD, amounting to
management tools over a continuous variation of initial conditions.
These depend on only three parameters: initial purity, entanglement
and excitation. Knowledge or control of initial phases is not
needed.
\end{abstract}

%\pacs{03.65.Ud, 03.65.Yz, 42.50.Pq, 75.10.Jm}

\maketitle

%+++++++++++++++++++++++++++++
\section{Introduction}
A continuously interacting background environment tends to destroy
quantum state coherence. Quantum non-separability (entanglement),
which plays a key role in various quantum computing and quantum
communication proposals \cite{ NC-Preskill}, is the best-known
example of non-local coherence. Regarding separability, it is known
that there is set of finite measure of separable states. If the
steady state solution of a master equation for an initially
entangled state lies on the interior of this set, then the
transition from an entangled to a separable state has to occur after
only a finite time of evolution. A version of these facts is
discussed in specific terms in connection with an event which is
called ESD (early-stage decoherence or more commonly entanglement
sudden death) by Yu and Eberly \cite{Yu-Eberly07} and by Al Qasimi
and James \cite{AlQasimi-James}. The inevitability of ESD for a
class of initial states does not, however, make available the actual
time when ESD will occur. Knowledge of the ESD time allows one, in
at least some situations, to undertake preparations to avoid or
delay its onset. In the following we address this question.

Studies have been made to examine the ways in which ESD depends on
dissipation mechanisms (e.g., amplitude damping \cite{amplitude},
phase damping \cite{phase}, and others \cite{other}), and/or on
interaction structures (e.g., interacting \cite{QQI} or
non-interacting \cite{NQQI} entangled partners, common \cite{common}
or independent \cite{independent} reservoirs), etc. Most of these
factors are difficult to control in practice.  In this Letter we
extend the study of ESD control in a straightforward way by
systematically deriving connections between ESD and initial
conditions, about which only isolated facts are noted, e.g., in the
study of ESD for Werner-like \cite{Ann-Jaeger} and X-type
\cite{Ali-etal} of initial states. We focus on the consequences of
the initial preparation of entanglement, purity, and double
excitation probability. Their initial values may be critical for an
experiment's design and also for avoiding noise-induced ESD during
the experimental time interval, but these initial values are not
always easy to establish precisely. However, for the generic case of
exposure to amplitude noise, one can identify initial-value phase
spaces which contain finite zones of safe preparation. That is,
within the well-defined boundaries of these zones the state
evolution is guaranteed not to be ESD-susceptible. To our knowledge,
this is the first indication that such phase boundaries exist for
any specific dissipation mechanism, and we provide formulas for the
boundary curves.

Additionally, analysis of the ESD-susceptible phases allows us to
determine another feature of ESD that has remained generically
unknown, namely the precise time when ESD will occur, if it is to
occur at all. The ESD onset time is found to be very sensitive to
the initial purity in the vicinity of critical purity points.
Especially for initial states with very high or low entanglement, we
show that a small amount of state impurity will accelerate the ESD
process dramatically. Among potential advantages in practice, the
conditions we have found allow protection from ESD and may be useful
in guiding experiments toward preparation of states that are immune
to ESD or have relatively longer ESD times.

\section{Prepared Initial State}

As an example we can consider the not quite perfect preparation of
Bell states $|\Phi^+\rangle$ and $|\Phi^-\rangle$, which are known
to be vulnerable to ESD in any combination. We suppose that a
quantum computation task aims to use a generic combination such as
\begin{equation}
|\Phi_{AB}\rangle =\cos \theta|e\rangle_{A}|e\rangle_{B}
+\sin\theta|g\rangle_{A}|g\rangle_{B},  \label{targetstate}
\end{equation}
where $|e\rangle_{A}$ and $|g\rangle_{A}$ are orthogonal states of
qubit $A$, etc., and may be interpreted as excited and ground
states. In this state the degree of entanglement between the two
qubits, as measured by concurrence \cite{Wootters}, is $C=\sin
2\theta$.

In reality, in preparation, a state will deviate from the ideal
target state. To sketch how this might occur we can begin with the
state after its preparation. It is coupled to an environment that
will cause post-preparation decoherence, and also still entangled
with marginal entities denoted by $M$, i.e., forces, fields and
objects that may participate in the preparation phase but then cease
interaction (see \cite{Qian-Eberly10, Brazil}). Thus we write:
\begin{eqnarray}
|\Phi (0)\rangle &=& \Big[\cos \theta
|e\rangle_{A}|e\rangle_{B}|m_{1}\rangle
\notag \\
& + & \sin \theta |g\rangle_{A}|g\rangle_{B}|m_{2}\rangle
\Big]\otimes |\phi_{0}\rangle_{a}|\phi_{0}\rangle_{b}.
\label{initial}
\end{eqnarray}
Here $\theta$ determines the degree of excitation of the two-party
state via $\cos ^{2}\theta =\rho_{11}$, $\sin ^{2}\theta
=\rho_{44}$, and $|\phi_{0}\rangle_{a}$ and $|\phi_{0}\rangle_{b}$
are the normalized initial states (usually ground states) of
environmental reservoirs $a$ and $b$ respectively, and
$|m_{1}\rangle $ and $|m_{2}\rangle$ are normalized states of the
marginal system $M$. Imperfect control of the preparation leads to a
possibly mixed rather than pure initial state via the relation $\sin
\theta \cos \theta \langle m_{2}|m_{1}\rangle =\rho_{14} \le
\sqrt{\rho_{11}\rho_{44}}$. Here $\rho_{ij}$ are the matrix elements
of the initial two-qubit reduced density matrix
\begin{equation}
\rho_{AB}(0)=\left(
\begin{array}{cccc}
\rho_{11} & 0 & 0 & \rho_{14} \\
0 & 0 & 0 & 0 \\
0 & 0 & 0 & 0 \\
\rho_{41} & 0 & 0 & \rho_{44}
\end{array}
\right).  \label{initial mixed}
\end{equation}

\section{Entanglement Evolution}

We next consider the dynamics of the two-qubit entanglement to see
how it is affected by the initial state specification, and expose
each qubit to the same amplitude damping process \cite{NC-Preskill}:
\begin{eqnarray}
|e\rangle_{A}|\phi_{0}\rangle_{a} &\longrightarrow &\sqrt{q_{a}}
|e\rangle_{A}|\phi_{0}\rangle_{a} +
\sqrt{p_{a}}|g\rangle_{A}|\phi_{1}
\rangle_{a}  \notag  \label{channel} \\
|g\rangle_{A}|\phi_{0}\rangle_{a} &\longrightarrow &
|g\rangle_{A}|\phi_{0}\rangle_{a},
\end{eqnarray}
where $|\phi_{1}\rangle_{a}$ is another environmental reservoir
state with $_{a}\langle \phi_{1}|\phi_{0}\rangle_{a}=0$, and $q_{a}$
is the probability that qubit $A$ remains in its excited state. For
the simplest illustration we assume exponential decay and take
$q_a(t) = \mathrm{exp}(-\Gamma t) \equiv \mathrm{exp}(-\tau)$. Then
the excitation-transfer probability $p_a = 1-q_a$ will grow from $0$
to $1$ irreversibly with increasing dimensionless time $\tau$. Qubit
$B$ and reservoir $b$ are characterized similarly as (\ref{channel})
with decay probability $q_b$. Of course this is more specialized
than needed - the rates of change $\Gamma$ need not be the same for
both qubits, and the amplitude channel can also be used to
characterize non-dissipative evolutions such as produced by the XY
spin interaction \cite{XY} or the Jaynes-Cummings interaction
\cite{JC}.

\begin{figure}[t!]
\includegraphics[width=4.2cm]{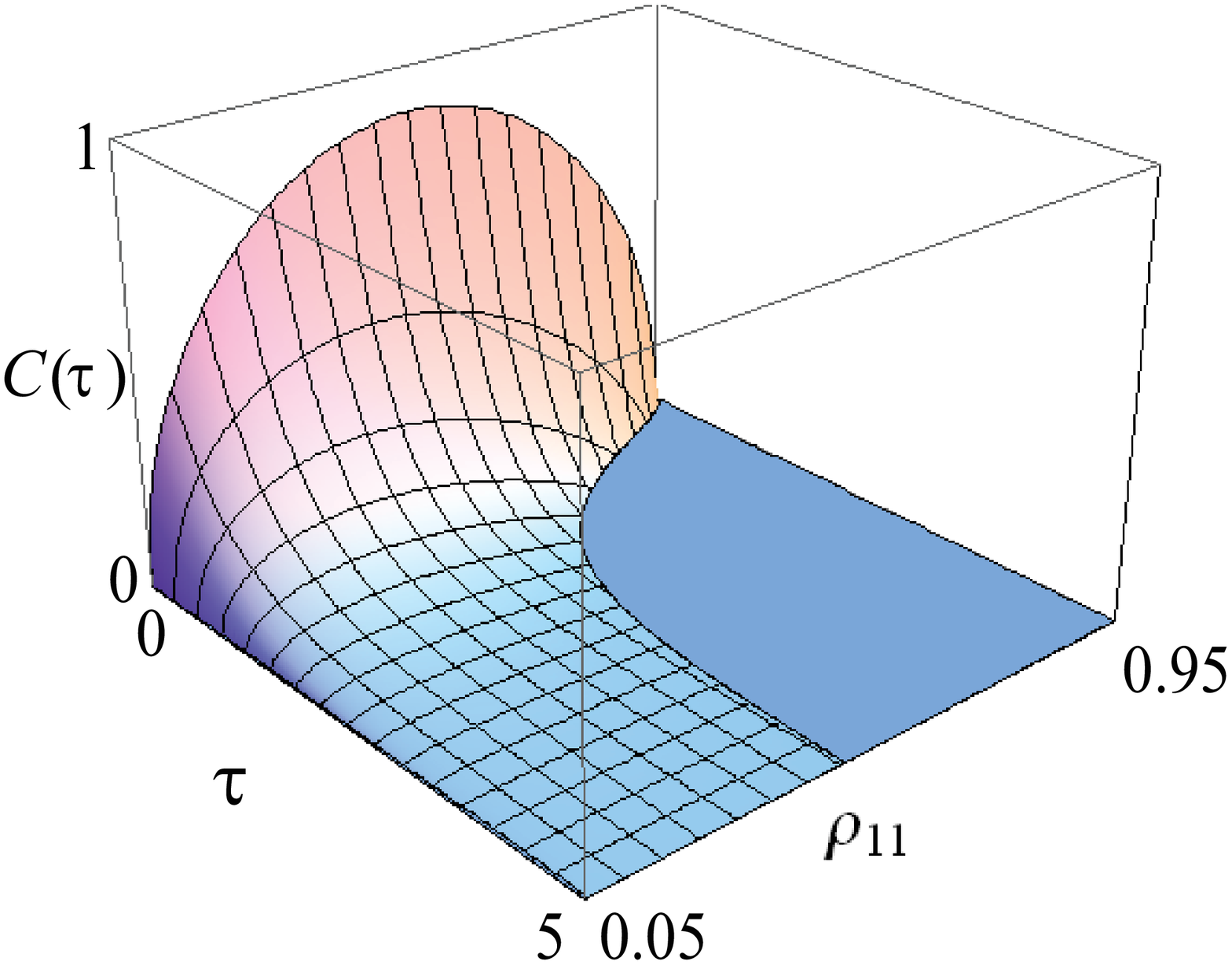}
\includegraphics[width=4.2cm]{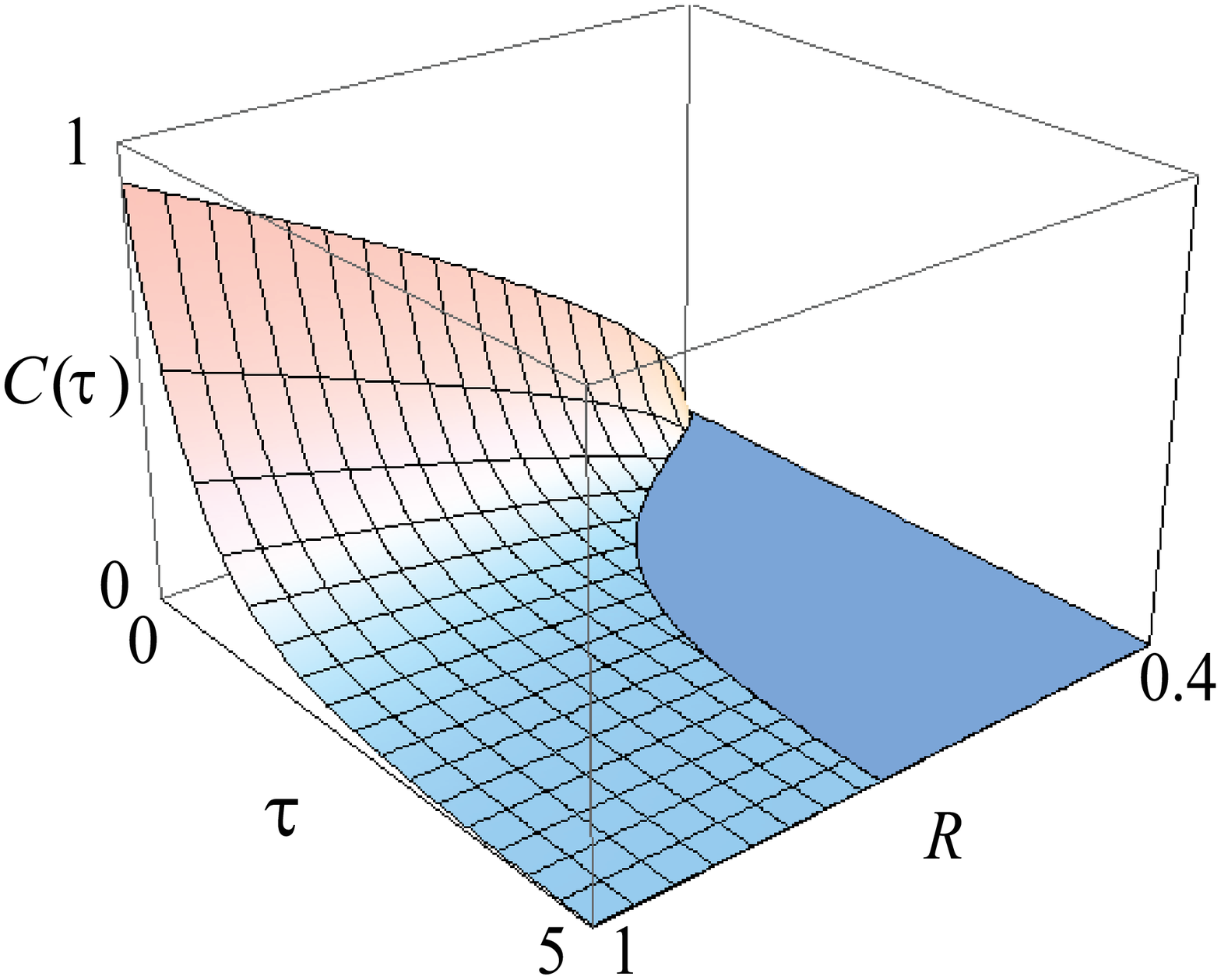}
\caption{Time-dependent surfaces showing entanglement evolution
parameterized in two ways: (left) $C\ vs. \tau$ for a range of
prepared double excitation values, and fixed purity; and (right) $C\
vs. \tau$ for a range of purities for fixed double excitation. The
symbols are defined in the text. In both plots the solid-color
base-plane shows the region where $C=0$, i.e., where ESD has already
occurred. It is clear that ESD-free regions are accessible, but the
extreme time-sensitivity of the ESD boundaries is not obvious here.}
\label{C vs t}
\end{figure}

Amplitude damping takes the initial state (\ref{initial}) to the
time-dependent state
\begin{eqnarray}
|\Phi (t)\rangle  &=&\cos \theta \sqrt{q_{a}q_{b}}|ee\rangle |\phi
_{0}\rangle _{a}|\phi _{0}\rangle _{b}|m_{1}(t)\rangle   \notag \\
&+&\cos \theta \sqrt{q_{a}p_{b}}|eg\rangle |\phi _{0}\rangle
_{a}|\phi
_{1}\rangle _{b}|m_{1}(t)\rangle   \notag \\
&+&\cos \theta \sqrt{p_{a}q_{b}}|ge\rangle |\phi _{1}\rangle
_{a}|\phi
_{0}\rangle _{b}|m_{1}(t)\rangle   \notag \\
&+&\cos \theta \sqrt{p_{a}p_{b}}|gg\rangle |\phi _{1}\rangle
_{a}|\phi
_{1}\rangle _{b}|m_{1}(t)\rangle   \notag \\
&+&\sin \theta |gg\rangle |\phi _{0}\rangle _{a}|\phi _{0}\rangle
_{b}|m_{2}(t)\rangle.
\end{eqnarray}
Here because of the fact that the marginal entity has ceased
interaction with the qubit system after $t=0$, the time dependent
marginal states preserve the overlap relation $\langle
m_{2}(t)|m_{1}(t)\rangle =\langle m_{2}|m_{1}\rangle $. Then the
two-qubit time-dependent reduced density matrix is obtained by
tracing off the reservoir and marginal states, yielding:

\begin{eqnarray}  \label{rhot}
&\rho_{AB}(t) = \\
&\left(
\begin{array}{cccc}
\rho_{11}q_a q_b & 0 & 0 & \rho_{14}\sqrt{q_a q_b} \\
0 & \rho_{11}q_{a}p_{b} & 0 & 0 \\
0 & 0 & \rho_{11}p_{a}q_{b} & 0 \\
\rho_{41}\sqrt{q_a q_b} & 0 & 0 & \rho_{11}p_a p_b +\rho_{44}
\end{array}
\right). \notag
\end{eqnarray}
The two-party entanglement (concurrence \cite{Wootters}) obeys the
X-state formula \cite{Yu-EberlyQIC}:
\begin{eqnarray}  \label{C(t)}
C(\tau) &=& 2\sqrt{q_{a}(\tau) q_{b}(\tau)}  \notag \\
& \times & max \{0,\ |\rho_{14}|-\rho_{11} \sqrt{p_{a}(\tau)
p_{b}(\tau)}\},
\end{eqnarray}
which is graphed in two ways in Fig. \ref{C vs t}.

Three parameters associated with the state mentioned above, that are
presumably under good if not perfect preparation control
\cite{Almeida-etalSci}, will now be selected to guide examination of
evolution. These parameters are $C_0,\ \rho_{11}$ and $R$, namely
the initial entanglement (concurrence), initial excited state
population, and a normalized value of purity. The true initial
purity is given by
\begin{eqnarray}  \label{true purity}
P &=& 1 - 2(\rho_{11}\rho_{44} - |\rho_{14}|^2)  \notag \\
&=& 1 - 2\rho_{11}(1-\rho_{11}) + C_0^2/2,
\end{eqnarray}
and the normalized version is defined as
\begin{equation}
R = \sqrt{2P - 1},
\end{equation}
which takes values conveniently between 0 and 1.

\section{ESD Susceptibility Phase Diagrams}

The plots in Fig. \ref{C vs t} are only for a specific category of
two-party states, but are systematically exhaustive in the sense
that any arbitrary combination of the $\Phi^{\pm}$ Bell states is
included, as well as any initial purity, excitation strength, and
concurrence. It turns out that their ESD boundaries are directly
analyzable, as follows.

First we use (\ref{rhot}) and (\ref{C(t)}) at $\tau = 0$, where $q =
1$ and $p = 0$, to characterize the entire physical domain in terms
of $\rho_{11}$ and $C_0$. Non-negativity of $\rho$ requires the
value of $\rho_{11}$ to be bounded between $\rho_{11}^{\max}$ and
$\rho_{11}^{\min}$, i.e., $\left[1 \pm \sqrt{1-C_0^{2}}\right]/2$,
for any fixed $C_0$. From the fact that $C_0 \leq 1$, one
immediately has
\begin{equation}
\rho_{11}^{\min} \leq C_0/2 \leq \rho_{11}^{\max}.
\end{equation}

\begin{figure}[!t]
\includegraphics[width=5cm]{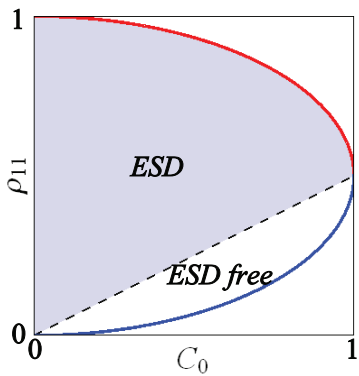}
\caption{Entanglement phase diagrams of the initial two-qubit state
in terms of the initial concurrence $C_0$ versus the double
excitation number $\rho_{11}$. The ESD phase (shaded area) and
ESD-free phase (unshaded area) are separated by the dashed critical
line $\rho_{11} = C_0/2$.} \label{PhaseDiagram1}
\end{figure}

Within this domain, as $\tau \rightarrow \infty $, both $p_a$ and
$p_b$ and their product grow from $0$ to $1$. Then it is easy to
identify within the physical domain the boundaries between ESD and
non-ESD phases. The condition $C(\tau) = 0$ tells us that ESD occurs
wherever
\begin{equation}  \label{first}
\sqrt{p_{a} p_{b}} \ge \frac{|\rho_{14}|}{\rho_{11}} =
\frac{C_0}{2\rho_{11}}.
\end{equation}
From the fact that $\sqrt{p_{a}p_{b}}\leq 1$, one immediately notes
that ESD is restricted to the domain where the initial double
excitation number is greater than half of the initial concurrence,
i.e., $C_0/2 < \rho_{11} \leq \rho_{11}^{\max}$. The opposite region
$\rho_{11}^{\min} \leq \rho_{11} \leq C_0/2$ is ESD-free. This leads
to the phase diagram in Fig. \ref{PhaseDiagram1}, which locates the
ESD and ESD-free phases in terms of initial concurrence $C_0$ and
the double excitation probability $\rho_{11}$. The solid (in color,
red and blue) boundary lines represent the values $\rho_{11}^{\max}$
and $\rho_{11}^{\min}$ as a function of $C_0$. The dashed line
$\rho_{11} = C_0/2$ is the critical phase boundary where the state
crosses from the ESD-free phase to the ESD-inevitable phase.

An important implication of this analysis is that one needs to
prepare the state with low double excitation, $\rho_{11} \leq C_0/2
\leq 1/2$, in order to avoid ESD. Therefore we will focus on this
region of greatest interest and assume $\rho_{11} \leq 1/2$ in the
following discussion.

\begin{figure}[!t]
\includegraphics[width=5cm]{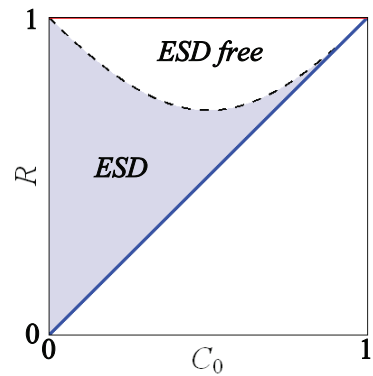}
\caption{Entanglement phase diagram in terms of the initial
concurrence $C_0$ versus the initial normalized purity $R$. The ESD
phase (shaded area) and ESD-free phase (unshaded area) are separated
by the dashed critical line $R_c = \sqrt{2C_0^{2} - 2C_0 + 1}$.}
\label{PhaseDiagram2}
\end{figure}

We now go a further step by employing the fact that the initial
concurrence $C_0$ and the double excitation probability $\rho_{11}$
are simply related to the initial state purity $P$ as in (\ref{true
purity}). Then the ESD-onset condition (\ref{first}) can be
rewritten in terms of the initial concurrence and normalized purity
$R$ as follows
\begin{equation}
\sqrt{p_{a} p_{b}} \ge \frac{C_0}{1-\sqrt{R^{2}-C_0^{2}}},
\label{second}
\end{equation}
where the non-negativity of the initial density matrix now requires
that the normalized purity $R$ obeys $R_{\min} = C_0$ and $R_{\max}
= 1$ for any fixed $C_0$. Fig. \ref{PhaseDiagram2} locates the ESD
and ESD-free phases in $C_0$-$R$ space. The (in color, red and blue)
solid lines represent the values $R_{\max}$ and $R_{\min}$ as
functions of $C_0$. The dashed line is defined by
\begin{equation}  \label{Rc}
R_c = \sqrt{2C_0^{2} - 2C_0 + 1}
\end{equation}
which specifies the critical purity, as a function of $C_0$, where
the state crosses from the ESD-free phase to the ESD-inevitable
phase.

One sees that the states with very high or very low initial
concurrences are quite fragile and require very high purity in order
to avoid ESD. States with initial concurrence around $0.5$ are the
most robust, because they tolerate the widest range $[\sqrt{1/2},1]$
of variation of $R$ or, in other words, have higher tolerance of
preparation errors (permitting lower initial purity) before breaking
down to be ESD-susceptible. This may have application in preparing
ESD-free states.

%+++++++++++++++++++++++++++++
\section{ESD Onset Time}

Although the phase boundaries are now located, there remains a
question of near-boundary values of concurrence. In the
ESD-inevitable regions the onset of ESD is not instantaneous so
temporary preservation of entanglement can be obtained if the
parameter region close enough to the phase boundary for ESD onset
can be well defined.

\begin{figure}[!h]
\includegraphics[width=5cm]{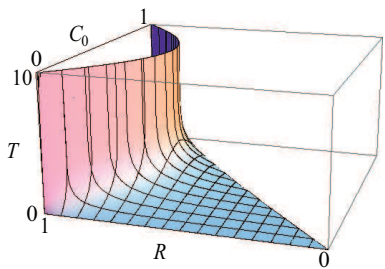}
\caption{Entanglement sudden death time as a function of the initial
entanglement $C_0$ and initial normalized purity $R$. The steepness
of the boundary indicates existence of a narrow zone of ``safe"
state preparation, even though susceptible to ESD.}
\label{tESDsurface}
\end{figure}

Within an ESD-susceptible zone we denote by $T$ the time when ESD
occurs and entanglement permanently vanishes, i.e., the ESD onset
time. That is, $T$ is the value of $\tau$ when relation
(\ref{second}) is an equality. With $p_{a} = p_{b} = 1-e^{-T}$, the
onset time is then given by
\begin{equation}
T = \ln \left[\frac{1-\sqrt{ R^{2}-C_0^{2}}}{1- C_0 -\sqrt{
R^{2}-C_0^{2}}}\right],  \label{ESDtime}
\end{equation}
which is displayed in Fig. \ref{tESDsurface} as a function of both
$R$ and $C_0$. One immediately sees from Fig. \ref{tESDsurface} that
for any fixed initial concurrence $C_0 \in [0,1]$, the ESD time is
very sensitive to the initial purity in the vicinity of its critical
values $R_{c}$. Especially for the states having very high or very
low initial concurrences, a small departure of the initial state
from $R=1$ will accelerate the ESD process dramatically.

\begin{figure}[h!]
\includegraphics[width=5cm]{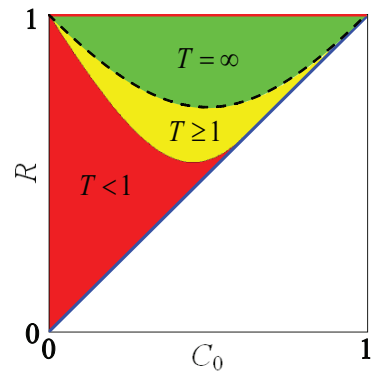}
\caption{ESD occurrence is identified here by time to onset of ESD.
The graph can be read as a traffic signal, where green is a safe
zone with no possibility of ESD, and red is the no-go zone where ESD
is sure to occur quickly (within one lifetime of the dissipative
process or less), and the yellow zone is one where ESD is certain to
occur but definitely deferred, i.e., ESD cannot occur before one
lifetime, and so permits informed state preparation with the initial
entanglement and purity values within its borders.} \label{tESD-RYG}
\end{figure}

An important practical point is that we can talk about three zones
with clear meaning. One is the zone of infinite onset time, where
ESD cannot occur, and one is the zone where ESD occurs inevitably
and very rapidly. A more interesting zone is the one where the onset
time for ESD is delayed in a definite way. A practical issue could
be that we can tolerate one dissipative lifetime of decay, but only
if we can be assured that ESD will not occur. This means that we
need to know the tolerable equivalent parameter range for the
initial preparation. To illustrate this case, we impose on Fig.
\ref{PhaseDiagram2} another boundary line, located at $T = 1$, and
show the result in Fig. \ref{tESD-RYG}. The yellow zone shows how
the parameter range available to the preparation process can be
expanded into the ESD-susceptible phase, tolerating some
entanglement dissipation while remaining certain to avoid ESD up to
the predetermined time $T = 1$.

\section{Summary}
In summary, we have reported new results bearing on the behavior of
non-local decoherence and its potential for being managed or even
controlled. As is well known and demonstrated \cite{Almeida-etalSci,
Eberly-Yu07} ESD exists, i.e., decoherence processes can drive
prepared entanglement to zero at the same time that easily monitored
local coherences and fidelity remain non-zero. Its inverse process,
sudden birth of entanglement \cite{ESB}, is equally interesting and
is the subject of a later report.

Until now there have been no rules of thumb or intuitive guides
giving reliable information about the likely occurrence or
non-occurrence of ESD even in the simplest instances of entanglement
and under the simplest of decoherence mechanisms. We have considered
a Bell superposition state as practically the simplest possible
entanglement scenario, and have subjected it to amplitude damping, a
well-understood decoherence mechanism. Similar steps have been taken
previously but only for specific initial density matrices generated
\emph{ad hoc} for demonstration purposes.

In contrast, our findings amount to a first step toward defining a
set of management tools that permit the range and extent of
entanglement decoherence to be bounded and controlled. These tools,
i.e., the values of initial purity, entanglement and excitation, are
based only on knowledge of parameters that are presumably under good
control during any specific entangled state preparation. More
importantly, the zone-control nature of these management tools,
i.e., the fact that ESD susceptibility is determined over separate
zones of a continuous variation of initial conditions, allows a
well-defined amount of margin for errors in practical imperfect
preparations. Notably, knowledge or control of initial phases is not
needed.

\section{Acknowledgement}
We acknowledge partial financial support from the following
agencies: DARPA HR0011-09-1-0008, ARO W911NF-09-1-0385, NSF
PHY-0855701.

\end{document}